\documentclass[aps,showpacs,preprint,pre,superscriptaddress]{revtex4}
\usepackage{graphicx}
\input amssym.tex

\begin{document}
\title{Encircling an Exceptional Point}
\author{C.~Dembowski}
\affiliation{Institut f{\"u}r Kernphysik, Technische Universit{\"a}t
Darmstadt, D-64289 Darmstadt, Germany}
\author{B.~Dietz}
\affiliation{Institut f{\"u}r Kernphysik, Technische Universit{\"a}t
Darmstadt, D-64289 Darmstadt, Germany}
\author{H.-D.~Gr{\"a}f}
\affiliation{Institut f{\"u}r Kernphysik, Technische Universit{\"a}t
Darmstadt, D-64289 Darmstadt, Germany}
\author{H.L.~Harney}
\affiliation{Max-Planck-Institut f{\"u}r Kernphysik, D-69029
Heidelberg, Germany}
\author{A.~Heine}
\affiliation{Institut f{\"u}r Kernphysik, Technische Universit{\"a}t
Darmstadt, D-64289 Darmstadt, Germany}
\author{W.~D.~Heiss}
\affiliation{Department of Physics, University of Stellenbosch, 7602
Matieland, South Africa}
\author{A.~Richter}
\affiliation{Institut f{\"u}r
Kernphysik, Technische Universit{\"a}t Darmstadt, D-64289 Darmstadt,
Germany}
\date{\today}

\begin{abstract} We calculate analytically the geometric phases that
the eigenvectors of a parametric dissipative two-state system described by a
complex symmetric Hamiltonian pick up
when an exceptional point (EP) is encircled. An EP is a parameter
setting where the two eigenvalues and the corresponding eigenvectors of
the Hamiltonian coalesce. We show that it can be
encircled on a path along which the eigenvectors remain approximately
real and discuss a microwave cavity experiment, where such an encircling of
an EP was realized. Since the wavefunctions remain
approximately real, they could be reconstructed
from the nodal lines of the recorded spatial intensity distributions
of the electric fields inside the resonator. We measured the geometric
phases that occur when an EP is encircled four times and thus confirmed
that for our system an EP is a branch point of fourth order.
\end{abstract}
 \maketitle

\section{Introduction} \label{sec:1} Since Berry's pioneering work
\cite{BerryGP} geometric phases, i.e. contributions to a quantum
system's phase which depend only on the geometry of the path traversed
by the system in its parameter space, have been the focus of intense
theoretical and experimental research. The majority of the theoretical
works discuss various generalizations of Berry's original paper, see
e.g. \cite{GPBuch,BerryDennis,Manini1,Manini2} and for a very early
work \cite{Voigt}, and investigate the appearance of geometric phases
in systems with complex eigenfunctions, as e.g. open or dissipative
systems \cite{GPOpen1,GPOpen2}. Most experimental works observe
geometric phases by tracing the pattern of nodal lines of a
wavefunction during adiabatic and cyclic processes
\cite{GPBuch,GPReview},  a technique suggested originally by Berry and
Wilkinson \cite{BerryGPRechnung}.

The dissipative nature of a system is commonly suppressed or neglected
in experiments (see e.g.~\cite{Lauber:94}). This was not the case in
the recently reported observation of a so-called Exceptional Point (EP)
in a microwave cavity experiment \cite{DemboEP1}. Such an EP, i.e. the
coalescence of two levels of a quantum system, occurs only in
dissipative systems, where it is associated with crossings and avoided
crossings of the eigenvalues \cite{Brentano,EPTheo,EPTheoNeu}. One of
the key features of an EP is the appearance of a geometric phase
\cite{DemboEP1,EPTheo,EPTheoNeu,Pancha} when it is encircled in
parameter space. If the EP is isolated, in its vicinity the dynamics is
predominantly determined by the two states corresponding to the
resonances, which coalesce at the EP. There, our system may be modelled
by a two-dimensional non-Hermitian, symmetric matrix. Such systems have
been analyzed in \cite{EPTheoNeu} and in \cite{BerryDennis}, where one
can find a complete and very detailed treatise on the essential
features of the eigenvalues and -vectors of parameter dependent
two-dimensional matrices associated with the singularities.

For two-state systems described by a complex symmetric Hamiltonian, the
geometric phase associated with an
EP \cite{DemboEP1,EPTheo} differs from that associated with a Diabolic
Point (DP) \cite{BerryGPRechnung}, a simple degeneracy between two
levels. The only way to determine a geometric phase with our experimental
setup \cite{DemboEP1} is by recording the change of the pattern of nodal lines
when encircling an EP in parameter space. In our microwave cavity experiments
we can only measure the intensity distribution of the electric field,
that is, the absolute value of the complex eigenfunctions. However,
a premise for the mere existence of a pattern of nodal lines is,
that the eigenfunctions remain real throughout the cyclic process.
Accordingly, the question
arises, whether Berry's reconstruction technique can be applied to the
complex eigenfunctions of the dissipative microwave resonator discussed in
\cite{DemboEP1,DemboEP2}.

In the present work we first focus on a parametric
two-state model adequate for the simulation of our experiment with the
dissipative microwave cavity. In doing so, we restrict ourselves
to the analysis
of those properties of the eigenvalues and eigenvectors, which are observable
using our experimental setup. For more detailed information one
might consult
\cite{BerryDennis}. Accordingly, we calculate the geometric phase
that occurs when an EP is encircled.
Moreover, we show that for this model a path around the
EP exists
along which the eigenvectors are approximately real, that is, have an
imaginary part being negligible compared to its real part. Indeed,
as is explained in more detail below,
from the mere fact, that we observe nodal lines rather than nodal points in our
experiment we may already conclude, that along the path
chosen in our experiment the eigenfunctions have exactly this property.
This then allows
us to employ Berry's reconstruction technique in a microwave cavity
experiment, where the development of the nodal line patterns with the
parameters is studied.

The paper is organized as follows: Using a two-state model adequate for the
simulation of our experiment we
analytically calculate in Sect.~\ref{sec:2} the geometric phases that
occur when an EP is encircled. By this calculation a path around
the EP is defined, along which the eigenvectors of the system remain
approximately real. In Sect.~\ref{sec:3} the actual microwave cavity
experiment is discussed. Using the same experimental techniques and
a setup similar to the one discussed in \cite{DemboEP1}, we here present
data from the encircling of a different EP. From the transformation properties
of the eigenfunctions and from the existence of a nodal line pattern
for every setting of the cavity's parameters we conclude that
the eigenfunctions remain approximately real in the
experiment while the EP is
encircled.  A conclusion is given in Sect.~\ref{sec:4}.

\section{Analytic Treatment of Encircling an EP}  \label{sec:2}  In this
section, we analyze the behavior of the eigenvectors of a two-state
quantum system when encircling an EP. As outlined in \cite{DemboEP1}, the
two-state Hamiltonian appropriate for the simulation of our system
in the vicinity of an isolated EP is written in terms of a two-dimensional
complex symmetric matrix. Accordingly, in the sequel
we restrict ourselves to such
two-state systems. In particular, we show that
a closed path exists along which the eigenvectors are approximately real
and that in accordance with \cite{BerryDennis,EPTheo} one needs four
turns around the EP in order to restore the
original situation. We use the same notation as in
\cite{HarneyHeiss,DemboEP2} and write the Hamiltonian of the two-state
system in the form
\begin{equation}
H=\pmatrix{E_1-i\gamma_1 & H_{12}        \cr
           H_{12}         & E_2-i\gamma_2
          }\, .
                       \label{HamMod}
\end{equation}
Here, the parameters $E_{1,2}$ and $\gamma_{1,2}$ are real and $H_{12}$
may be complex. Expression (\ref{HamMod}) is a complex symmetric
Hamiltonian. By defining
\begin{equation}
\mathcal{E}\equiv{{E_1+E_2}-i{(\gamma_1+\gamma_2)}\over 2}
\end{equation}
and
\begin{eqnarray}
e      &\equiv&(E_1-E_2)/2\, ,          \nonumber\\
\gamma &\equiv&(\gamma_1-\gamma_2)/2\, , \end{eqnarray} the
eigenvalues, $E_{1,2}$, of $H$ can be written as
\begin{equation}
E_{1,2}=\mathcal{E}\pm\sqrt{(e-i\gamma)^2+H_{12}^2}\, .
\end{equation}
Two-state systems described by such a Hamiltonian have been studied
both analytically and numerically in \cite{EPTheoNeu}. Moreover,
subtracting $\mathcal{E}$ from the diagonal elements of $H$ given in
\cite{BerryDennis} and performing a transformation of the type defined
in equation (3.6) of \cite{BerryDennis}, the Hamiltonian (\ref{HamMod})
can be brought to the form given in equation (6.2) of
\cite{BerryDennis} with G=0, whose eigenvalues and eigenvectors provide
the refractive indices and the associated polarization vectors of
dichroic, non-chiral crystals. Reference \cite{BerryDennis} provides a
very detailed description of such crystals at and around three types of
singularities, (called singular axes, C-points of circular
polarization, which in the absence of chirality coincide with the
singular axes, and L-lines of linear polarization) that may occur
dependent on the choice of the three parameters $e$, $\gamma$, and
$H_{12}$. In the following we will rederive those properties of the
eigenvalues and eigenvectors of the Hamiltonian (\ref{HamMod}), which
are observable using our experimental setup.

The complex eigenvalues coincide if the square root vanishes. Hence, at
the two EPs of $H$, one has the relation
\begin{equation}
H_{12} =\pm i(e-i\gamma )
\label{EPCond}
\end{equation}
between the parameters of the Hamiltonian. Furthermore, the parameter
$H_{12}$ is non-zero at an EP. Else, i.e. if $H_{12}$ vanishes, the
space of eigenvectors is two-dimensional (see Eq. (\ref{HamMod})), and
a degeneracy rather than an EP occurs.

Since we are interested in the behavior of the eigenvectors of $H$ in
the vicinity of an EP, we furthermore define the complex parameter
\begin{equation}
B\equiv{{e-i\gamma}\over H_{12}}\, ,
                    \label{DefB}
\end{equation}
which becomes
\begin{equation}
B^{\rm EP}=\pm i\,
\end{equation}
at an EP, i.e. when Eq.~(\ref{EPCond}) is fulfilled. The eigenvectors,
$|r_1\rangle$ and $|r_2\rangle$, of $H$  can then be written as
functions of $B$. Normalizing the left hand eigenvectors $\langle
l_k\vert$ and the right hand eigenvectors $\vert r_k\rangle$ in the
biorthogonal sense, they can be defined as
\begin{eqnarray}
\langle l_1| = \left( \cos\theta\, , \, \sin\theta\right)\, &,& \, |r_1\rangle =
\left(\begin{array}{c}
\cos\theta \\
\sin \theta
\end{array}\right) \nonumber\\
\langle l_2| = \left( -\sin\theta\, , \, \cos\theta\right)\, &,& \,
|r_2\rangle =
\left(\begin{array}{c}
- \sin\theta \\
\cos \theta
\end{array}\right)
\label{EigenvecMod}
\end{eqnarray}
where $\theta$ is defined by
\begin{eqnarray}
\tan\theta &\equiv&-B+\sqrt{B^2+1}\nonumber\\
           &= &-B+\sqrt{(B+i)}\sqrt{(B-i)}\, .
\label{DefTheta}
\end{eqnarray}
This choice of normalization of course defines the left and right
eigenvectors only up to an additional phase, which cancels out when
evaluating the absolute value of the eigenvectors. Hence, this
additional phase may not be observed with our experimental setup (see
Sect.~\ref{sec:3neu}).

When varying $B$ continuously along a closed curve around one of the
EPs, i.e. around $B^{\rm EP}=+i$ or $B^{\rm EP}=-i$, the phase  of
$B-B^{\rm EP}$ will change by $2\pi$. Accordingly, \emph{one} of the
square-root functions in the second line of Eq.~(\ref{DefTheta}),
namely that corresponding to $\sqrt{B-B^{\rm EP}}$, changes its sign,
whereas $\sqrt{B+B^{\rm EP}}$ will return to its original value as long
as exactly one EP is encircled. Hence, encircling an EP implies a
change from $\tan\theta$ to $\tan\theta_1$, where
\begin{eqnarray}
\tan\theta_1 &\equiv&-B-\sqrt{B^2+1}\nonumber\\
             &=&-\cot\theta\, ,
\end{eqnarray}
that is,
\begin{equation}
\theta_1=\theta\pm\frac{\pi}{2}\, .
\end{equation}
To compare the eigenvectors of $H$ before and after encircling an EP in
the $B$-plane, we
use the abbreviation
\begin{equation}
|1,2\rangle \equiv |r_{1,2}\rangle\Big|_{B_0}\, ,
\label{EigenvekAbkuerz}
\end{equation}
where $B_0$ denotes some value of $B$. Starting from this initial value
$B_0$ we track the development of the eigenvectors of $H$ when an EP is
encircled. Comparing the eigenvectors before and after encircling an
EP, i.e. inserting $\theta$ and $\theta_1=\theta+\frac{\pi}{2}$ into
Eq. (\ref{EigenvecMod}),  then yields the transformation scheme
\begin{equation}
\left\{\begin{array}{c}
       |1\rangle\\
       |2\rangle
      \end{array}
\right\} \circlearrowleft
\left\{\begin{array}{r}
        |2\rangle\\
        -|1\rangle
      \end{array}
\right\} \circlearrowleft
\left\{\begin{array}{c}
       -|1\rangle\\
       -|2\rangle
      \end{array}
\right\} \circlearrowleft
\left\{\begin{array}{c}
       -|2\rangle\\
       |1\rangle
      \end{array}
\right\} \circlearrowleft
\left\{\begin{array}{c}
       |1\rangle\\
       |2\rangle
      \end{array}
\right\}\, ,
              \label{PhasenA}
\end{equation}
while inserting $\theta$ and $\theta_1=\theta-\frac{\pi}{2}$ leads to
\begin{equation}
\left\{\begin{array}{c}
       |1\rangle\\
       |2\rangle
      \end{array}
\right\} \circlearrowright
\left\{\begin{array}{r}
        -|2\rangle\\
        |1\rangle
      \end{array}
\right\} \circlearrowright
\left\{\begin{array}{c}
       -|1\rangle\\
       -|2\rangle
      \end{array}
\right\} \circlearrowright
\left\{\begin{array}{c}
       |2\rangle\\
       -|1\rangle
      \end{array}
\right\} \circlearrowright
\left\{\begin{array}{c}
       |1\rangle\\
       |2\rangle
      \end{array}
\right\}\, .
              \label{PhasenB}
\end{equation} Both schemes have been observed experimentally
\cite{DemboEP1,DemboEP2} in a particularly shaped microwave billiard
(see Sect.~\ref{sec:3}) by changing the orientation of the closed loop
around the EP, as indicated by the symbols $\circlearrowleft$ and
$\circlearrowright$. Changing the orientation of the loop is equivalent
to following a given scheme backwards. In both cases, \emph{four} turns
around the EP are needed in order to restore the original situation
(see also \cite{BerryDennis,DemboEP1,EPTheo}). At this point we note,
that if $B$ changes continuously along a closed curve, which encircles
\emph{both} EPs, then the sign of \emph{both} square-root functions in
the second line of (\ref{DefTheta}) will change, that is,  each of the
eigenvectors $|1\rangle$ and $|2\rangle$ transforms into itself. If,
however, both EPs are encircled with opposite orientation by tracing
out an '8', then both eigenvectors will acquire an extra phase $\pi$,
that is a phase which coincides with Berry's phase for the encircling
of a diabolical point (see Sect.~\ref{sec:3neu}).

\begin{figure}[ht] \includegraphics[width=0.8\textwidth]{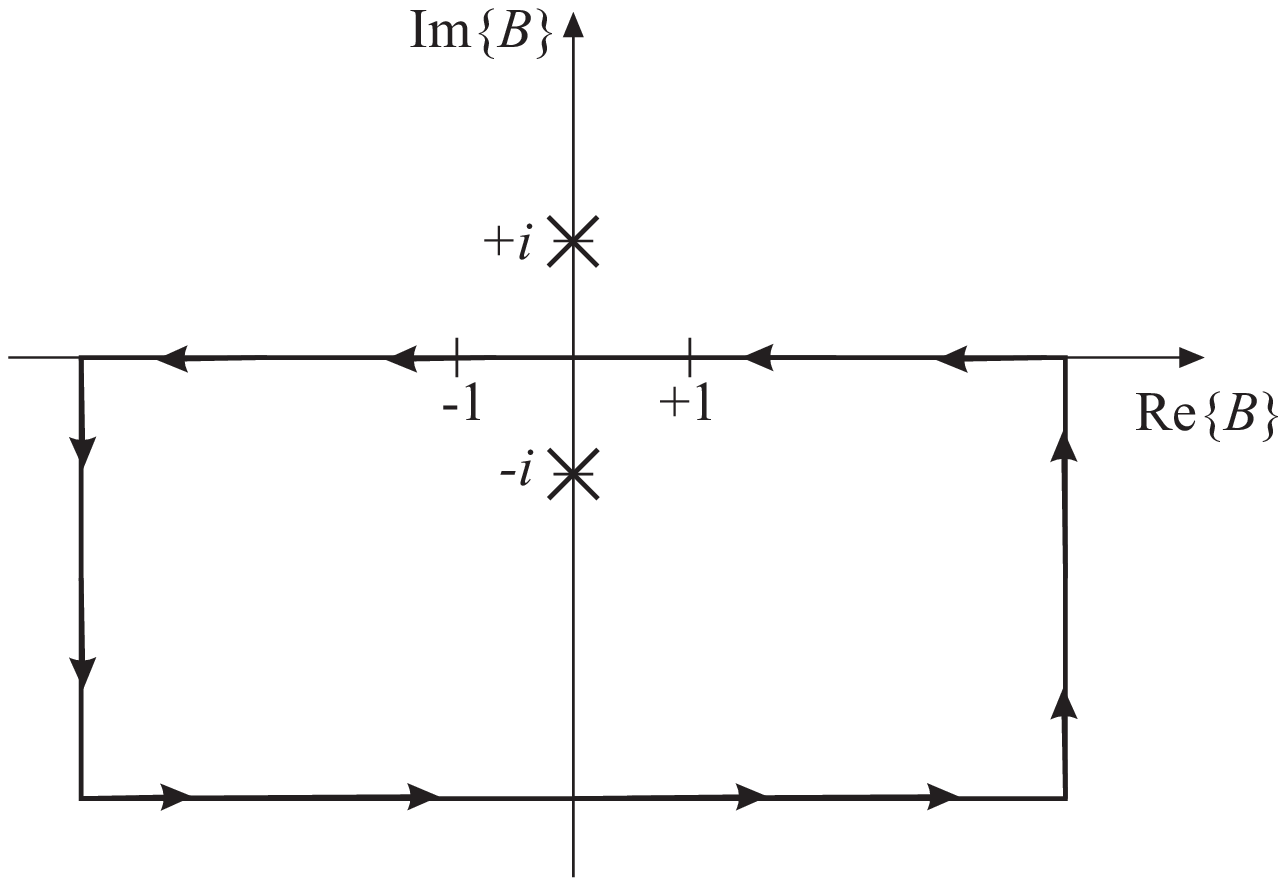}
\caption{A path in the complex $B$-plane (cf.~Eq.~(\ref{DefB}) in the
main text) surrounding an EP situated at $B^\mathrm{EP}=-i$. Along this
path the eigenfunctions of $H$ remain approximately real.}
\label{fig:1} \end{figure}

The resulting transformation schemes (Eqs.~(\ref{PhasenA}) and
(\ref{PhasenB})) are not surprising since the  eigenvectors
(\ref{EigenvecMod}) have a branch point of a fourth root at the EP.
This can be easily seen by noting that (i) the tangent in
Eq.~(\ref{DefTheta}) depends on $\sqrt{B\pm B^{\rm EP}}$ and (ii) an
additional square root is needed for the computation of the sine- and
cosine-functions in Eq.~(\ref{EigenvecMod}), from the tangent viz.
\begin{eqnarray}
\cos\theta &=&\frac{1}{\sqrt{1+\tan^2\theta}}\, ,\nonumber\\
\sin\theta &=&\frac{1}{\sqrt{1+\cot^2\theta}}\, .
\end{eqnarray}
If $B$ is real, the components of $|1\rangle$ and $|2\rangle$,
are real. Hence, if $B$ is sufficiently far from the EPs, i.e.
\begin{equation}
|B|\gg 1\, ,
         \label{LargeB}
\end{equation}
the eigenvectors are approximately real. Keeping only terms up to the
first order in $B^{-1}$ we obtain
\begin{eqnarray}
|r_1\rangle &\approx &\left(\begin{array}{c}
                                1        \\
                                (2B)^{-1}
                               \end{array}
                         \right)\, ,\nonumber\\
|r_2\rangle &\approx &\left(\begin{array}{c}
                                (2B)^{-1}\\
                                -1
                               \end{array}
                         \right)\, .
                         \label{EigenvecApprox}
\end{eqnarray}
Therefore the eigenfunctions are approximately real along the path
sketched in Fig. \ref{fig:1}, which either follows the real axis of the
complex $B$-plane or fulfills Eq.~(\ref{LargeB}). We note here that
along the lower part of the path sketched in Fig.~\ref{fig:1}, i.e. far
away from $B^\mathrm{EP}=-i$, the components of the eigenvectors depend
only linearly on $B^{-1}$. We therefore expect, that the experimentally
measured eigenvectors vary only slightly along this part of the path.
If the nodal line patterns are tracked along this path in an
experiment, they can be used to reconstruct the eigenstates according
to Berry and Wilkinson \cite{BerryGPRechnung}.

\section{Phase of the Eigenvectors}
\label{sec:3neu}

In the following section we will discuss why \emph{only} the
transformation schemes (\ref{PhasenA}) and (\ref{PhasenB}) are observed
in experiments (see \cite{DemboEP1} and Sect.~\ref{sec:3}). One can
redefine the eigenvectors such that
\begin{equation} \langle l_k| \rightarrow \langle \tilde{l}_k
| = e^{-i\phi} \langle l_k| \quad , \quad |r_k\rangle \rightarrow
|\tilde{r}_k \rangle = e^{i\phi} | r_k\rangle
\label{EigenvectorsWithPhase}
\end{equation}
This conserves the
biorthogonal normalization. If the phase $\phi$ is e.g. constructed
\cite{Rotter} so that after two loops around an EP it equals $\pi$ then
one obtains $|\tilde{r}_k\rangle \rightarrow |\tilde{r}_k\rangle$ after
two loops. This seems to disagree with the schemes
(\ref{PhasenA},\ref{PhasenB}) claiming that $|\tilde{r}_k\rangle
\rightarrow -|\tilde{r}_k\rangle$ after two loops. In the present
section we show that the experimental result (\ref{PhasenA}) for the
phase change of the eigenvectors (\ref{EigenvecMod}) remains unchanged
under the replacement (\ref{EigenvectorsWithPhase}). In the first
subsection the essence of the argument is presented in a rather general
and abstract way. In the second subsection the loops around an EP are
described in a more physical way. In the third subsection, we consider
eigenstates with an additional phase, as defined in
(\ref{EigenvectorsWithPhase}), and show, that the results of the second
subsection remain unchanged. In the last subsection we show that the
present arguments yield the well-known geometrical phase occurring when
a DP is encircled.

\subsection{Smoothest interpolation between the experimental pictures}
\label{sec:3neu.1}

In the experiment described in Sec. \ref{sec:3} and in
\cite{DemboEP1,DemboEP2} the development of the eigenvectors
(\ref{EigenvecMod}) is tracked. Their coefficients are analytical
functions of $B$ everywhere except at the EPs. Analytical functions are
arbitrarily often differentiable. In this sense they are the smoothest
possible functions.

Except for a parameter-independent phase, the vectors
(\ref{EigenvecMod}) are even the only analytical representation of the
eigenvectors, because a system of biorthogonal eigenvectors is well
defined up to an arbitrary phase $\phi =\phi(B)$ as introduced in
(\ref{EigenvectorsWithPhase}) with a complex $B$. The phase is real in
all the domain where one is allowed to choose the path $C$. Except for
the constant there is no analytical function which is real on some area
in the complex plane. Hence, multiplying (\ref{EigenvecMod}) with a
phase factor depending on $B$ yields non-analytical eigenvectors.

Instead of claiming that the experiment follows (\ref{EigenvecMod}),
one can therefore state that (\ref{EigenvecMod}) is the smoothest
interpolation between the experimental pictures of the eigenfunctions.
In this sense it is the simplest mathematical interpretation of the
sequence of wave functions presented in Fig. 3. According to the
argument of Ockham's razor \cite{Ockham} one cannot hope for anything
else.

In the next subsection, a physical process is discussed that allows to
explicitly follow a given eigenvector on a path encircling an EP. It
yields in Sect. \ref{sec:3neu.3} the result announced above.

\subsection{Parameter-dependent state}
\label{sec:3neu.2}

The physical process is the one introduced by Berry in \cite{BerryGP}
-- modified to the treatment of complex symmetric (instead of
hermitian) $H$ and to loops around an EP (instead of a DP).

Let $H=H(\vec{R})$ depend on a set $\vec{R}$ of parameters. The
eigenvectors $\langle l_k (\vec{R})|$, $| r_k (\vec{R})\rangle$ and
eigenvalues $E(\vec{R})$ depend on $\vec{R}$.

In a first step, we consider $\vec{R} = \vec{R}(t)$ as a function of
time. The state $|\psi(t)\rangle$ of the system is the solution of the
time dependent Schr{\"o}dinger equation
\begin{equation}
H(\vec{R}(t))|\psi(t)\rangle=i\frac{\partial}{\partial t}
|\psi(t)\rangle\, . \label{SchroedingerEq}
\end{equation}
At $t=0$ the
system shall be in the eigenstate $|r_n\rangle$, i.e.
\begin{equation}
|\psi(0)\rangle=|r_n(\vec{R}(0))\rangle\, . \label{EigenvStart}
\end{equation}
Expanding $|\psi(t)\rangle$ into the instantaneous
eigenstates $\vert r_n(\vec R(t))\rangle$ at time $t$ and assuming,
that the parameter $\vec R (t)$ is changed so slowly, that the
adiabatic approximation is applicable \cite{Nenciu}, we obtain
\begin{equation} |\psi(t)\rangle = \exp\left( -i
\int_0^tdt'E_n(\vec{R}(t'))\right) \exp\left(i\phi (\vec{R}(t'))\right)
|r_n(\vec{R}(t))\rangle\, . \label{PhaseIntegral}
\end{equation}
Hence,
the adiabatic approximation implies, that at each instant $t$ the state
$\vert\psi (t)\rangle$ is proportional to the instantaneous eigenstate
$\vert r_n(\vec R(t))\rangle$, if it is proportional to $\vert r_n(\vec
R(0))\rangle$ at time $t=0$. The dynamical phase $$-i
\int_0^tdt'E_n(\vec{R}(t'))$$ is well known. Of special interest is the
additional phase $\phi$ which is due to the motion in parameter space.
With the ansatz (\ref{PhaseIntegral}), Schr{\"o}dingers equation
(\ref{SchroedingerEq}) yields the equation
\begin{equation} \dot{\phi}
|r_n\rangle = i \frac{\partial}{\partial t} |r_n\rangle
\label{SchroedAlpha}
\end{equation}
for the phase $\phi$. Using
biorthogonality this gives
\begin{eqnarray}
\dot{\phi} &=& i \langle l_n| \frac{\partial}{\partial t} r_n\rangle\nonumber \\
&=& i \langle l_n| \vec{\nabla}_R r_n\rangle \dot{\vec{R}}\, .
\label{AlphaEq1}
\end{eqnarray}
The solution is
\begin{eqnarray}
\phi &=& i \int_0^t dt'\langle l_n| \vec{\nabla}_R r_n \rangle
\dot{\vec{R}} \nonumber \\
&=& i \int_C d\vec{R} \langle l_n| \vec{\nabla}_R r_n\rangle\, .
\label{AlphaEq2}
\end{eqnarray}
The last integral is a path integral in
$\vec{R}$-space. The parameters move along the path $C$ between time
zero and time $t$. This means that the time has only served to
parameterise the path. The last integral is independent of time and
should therefore be valid not only for an adiabatic process but also in
the present experimental context, where a continuous variation of the
parameters is considered. Let
\begin{equation} \vec{R}= \left(
\begin{array}{c} \mathrm{Re} B \\ \mathrm{Im} B \end{array} \right)
\label{DefB2}
\end{equation}
be the real and imaginary parts of B, and
let $\langle l_n |$, $|r_n\rangle$ be defined as in
(\ref{EigenvecMod}). Then (\ref{AlphaEq2}) is a path integral in the
complex plane of B, viz.
\begin{equation}
\phi = i \int_C dB \langle
l_n| \frac{d}{d B} r_n\rangle \, . \label{AlphaEq3}
\end{equation}
The
integrand is analytic everywhere except at the EPs. The function
$\langle l_n| \frac{d}{d B} r_n\rangle$ is multivalued -- it is defined
on a Riemannian surface rather than the complex plane. But on that
surface, it is analytic everywhere. Especially it is continuous on
every path $C$ that avoids the EPs. Along such a path one has
\begin{eqnarray} \langle l_n | \frac{d}{d B} r_n\rangle &=& \frac{1}{2}
\left( \langle \frac{d}{d B} l_n | r_n \rangle + \langle l_n |
\frac{d}{d B}
r_n\rangle \right)\nonumber\\
&=& \frac{1}{2} \frac{d}{d B} \langle l_n | r_n \rangle\nonumber\\
&=& 0\, .
\label{ScalarProdLR}
\end{eqnarray}
The first line of this equation holds because $\langle l_n|$ is just
the transpose of $|r_n\rangle$. The result is due to the biorthogonal
normalisation.

Hence, the choice (\ref{EigenvecMod}) for the eigenfunctions leads to
\begin{equation}
\phi (C) = 0 \label{AphaofC}
\end{equation}
if the
path $C$ does not cross an EP. In other words: Our choice of the
normalization implies that the total phase acquired by a state
$\vert\psi (t)\rangle$ when encircling an EP in parameter space is
obtained from the change of the eigenstates, that is, from the
transformation schemes (\ref{PhasenA},\ref{PhasenB}). As will be shown
in the next subsection, this result is independent of the phase
convention chosen for the eigenstates.

\subsection{Redefining the phase of the eigenstates}
\label{sec:3neu.3}
Let us use in Sec.~\ref{sec:3neu.2} the eigenstates
$|\tilde{r}_k\rangle$, $\langle \tilde{l}_k|$ of
\ref{EigenvectorsWithPhase} instead of $|r_k\rangle$, $\langle l_k|$.
The phase $\phi=\phi(\vec{R})$ shall be a function of the parameters.
Then, in analogy to Eq.~(\ref{PhaseIntegral}) we may write
\begin{eqnarray} |\psi(t)\rangle &=& \exp\left( -i
\int_0^tdt'E_n(\vec{R}(t'))\right)
\exp\left(i\beta (\vec{R}(t'))\right) |\tilde{r}_n(\vec{R}(t))\rangle\nonumber\\
&=&  \exp\left( -i \int_0^tdt'E_n(\vec{R}(t'))\right) \exp\left(i\beta
(\vec{R}(t)) + \phi(\vec{R}(t))\right) |r_n(\vec{R}(t))\rangle\,
,\label{PhaseIntegralBeta} \end{eqnarray}
where in the second line we
used the definition of $|\tilde{r}_n\rangle$, (see
Eq.~\ref{EigenvectorsWithPhase}). Hence, the total change of the phase
of $|\psi(t)\rangle$ is given as $(\beta(C) + \phi(C))$ plus the change
of $|r_n\rangle$. But, proceeding as in Sect.~\ref{sec:3neu.1} one
shows that $(\beta(C) + \phi(C))=0$, as long as the path does not cross
an EP. As a result, our choice of the normalization of the eigenstates
has the specific property, that all phase changes acquired by a state
$\vert\psi (t)\rangle$ when encircling an EP are solely obtained from
the transformation schemes (13,14) independent of the phase conventions
chosen for the eigenstates -- in agreement with the argument given in
Sect. \ref{sec:3neu.1}. For comparison we briefly discuss -- in the
following subsection -- the well-known phase that occurs when a
diabolic point is encircled.

\subsection{Encircling a Diabolic Point} \label{sec:3neu.4}
A diabolic
point is a degeneracy of the eigenvalues such that there are two
linearly independent eigenvectors. It is most easily obtained as a
degeneracy in a system described by a real, symmetric $H$. Let us set
in Eq.~(\ref{HamMod})
\begin{equation} \gamma_1 = 0 \quad , \quad
\gamma_2=0 \label{GammasZero}
\end{equation}
and
\begin{equation}
H_{12} = \omega= \mathrm{real}\, . \label{H12Zero}
\end{equation}
Then
the diabolic point occurs when $e=0$ and $H_{12}=0$. Encircling
corresponds to moving the vector $$\left( \begin{array}{c} e \\ H_{12}
\end{array}\right)$$ around the origin, e.g. with
\begin{equation}
e=\varrho \cos \xi \quad , \quad H_{12}=\varrho \sin \xi \quad , \quad
0 \leq \xi < 2 \pi\, . \label{Parametrization}
\end{equation}
This
yields the eigenvectors
\begin{equation}
|r_1\rangle =
\left( \begin{array}{c}
\cos (\xi /2)\\
\sin (\xi /2)
\end{array}
\right)
\label{Eigenvec1}
\end{equation}
and
\begin{equation}
|r_2\rangle =
\left(
\begin{array}{c}
\sin (\xi /2)\\
-\cos (\xi /2)
\end{array} \right)\, . \label{Eigenvec2}
\end{equation}
One sees that encircling a DP changes the sign of the eigenvectors --
this is Berry's phase \cite{BerryGP}.

Strictly speaking, the path (\ref{Parametrization}) cannot be represented as a
closed curve in the complex plane of $B$. According to (\ref{Parametrization}),
$B$ is real for all $\xi$ except $\xi =0,\pi$, where it goes to infinity.
In order to circumvent this difficulty one can add a small imaginary part to
$H_{12}$ and replace (\ref{H12Zero}) by
\begin{equation} H_{12} =
\omega + i \epsilon\, . \label{H12SmallImaginary}
\end{equation}
By virtue of (\ref{Parametrization}), the parameter
\begin{equation}
B=\frac{e}{\omega + i \epsilon} \label{BPath}
\end{equation}
then moves on a path that has the shape of a figure-eight and encircles
the EPs at $i$ and $-i$ in opposite directions.  We have convinced
ourselves that such a closed path changes the phase of the eigenvectors
by $\pi$ -- in agreement with Berry's phase.

The parameterisation in terms of $B$ is similar to the one discussed in
\cite{EPTheoNeu}. However, the authors of \cite{EPTheoNeu} did not exactly
specify the path chosen for encircling a DP. Note that a closed path
which encircles both EPs in the same sense does not change the phase of the
eigenvectors.

\section{Experiment}
\label{sec:3}

\begin{figure}[ht]
\includegraphics[width=0.8\textwidth]{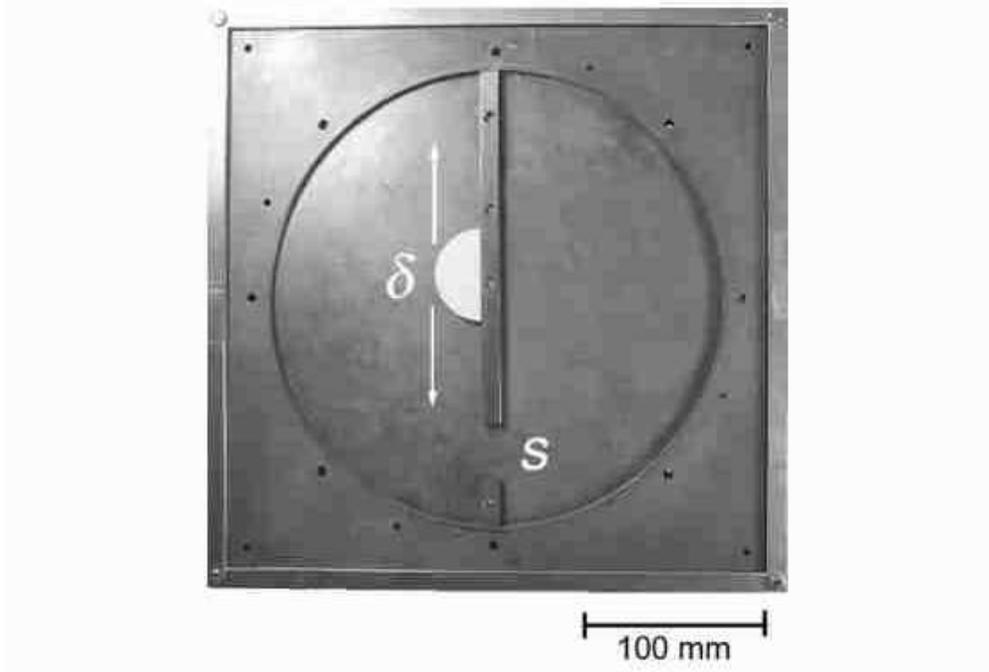} \caption{A
photograph of the opened microwave billiard employed for the
observation of EPs. A circular copper cavity is divided into two
semi-circular parts. The two parts are variably coupled by a slit of
width $s$. One of the semi-circular cavities can be perturbed by
adjusting the position $\delta$ of a teflon stub inside the resonator.}
\label{fig:2} \end{figure}

The geometric phases which occur when an EP is encircled have been
observed for the first time in the microwave cavity experiment
described in \cite{DemboEP1}. Flat microwave resonators as the one used
in \cite{DemboEP1} are commonly known as microwave billiards and form
one cornerstone for the experimental investigation of quantum chaotic
phenomena (for an overview see e.g. \cite{StoeckBuch,RichterBuch}).
They compose an analog computer that solves the Schr{\"o}dinger
equation for quantum billiards. The circular resonator employed in the
experiment \cite{DemboEP1} was  manufactured of copper and divided by a
conducting wall into two approximate half-circles. Figure \ref{fig:2}
shows a photograph of the cavity without its lid. An opening of length
$s$ in the wall couples the two semi-circular parts of the cavity. A
second adjustable parameter, called $\delta$, is given by the position
of a semi-circular Teflon stub in one part of the cavity.

In the vicinity of an EP \cite{DemboEP1}, we model the microwave
billiard through a two-state system as described by $H$ of
Eq.~(\ref{HamMod}). The connection between the parameters of $H$ and
the observables of the microwave cavity experiment, can be illustrated
directly for the uncoupled case which implies $s=0$~mm, that is
$H_{12}=0$, in Eq.~(\ref{HamMod}): The resonance frequency $f_1$ of a
mode in one semi-circular cavity corresponds to $E_1$ in
Eq.~(\ref{HamMod}) while the total width of it, $\Gamma_1$, corresponds
to $\gamma_1$. The resonance frequency  and total width of a mode in
the adjacent semi-circular cavity gives then $E_2$ and $\gamma_2$. The
situation is slightly more involved for $s>0$~mm, i.e. $H_{12}\neq 0$.
However, the the diagonal elements of $H$ are the same as in the
uncoupled case. Moreover, the coupling mechanism via the slit implies
that the off-diagonal elements of $H$ coincide, that is, $H$ is complex
symmetric. The resonance frequencies and the widths measured in the
experiment correspond to the real and the imaginary part of the
eigenvalues of $H$.

\begin{figure}[ht] \includegraphics[width=0.8\textwidth]{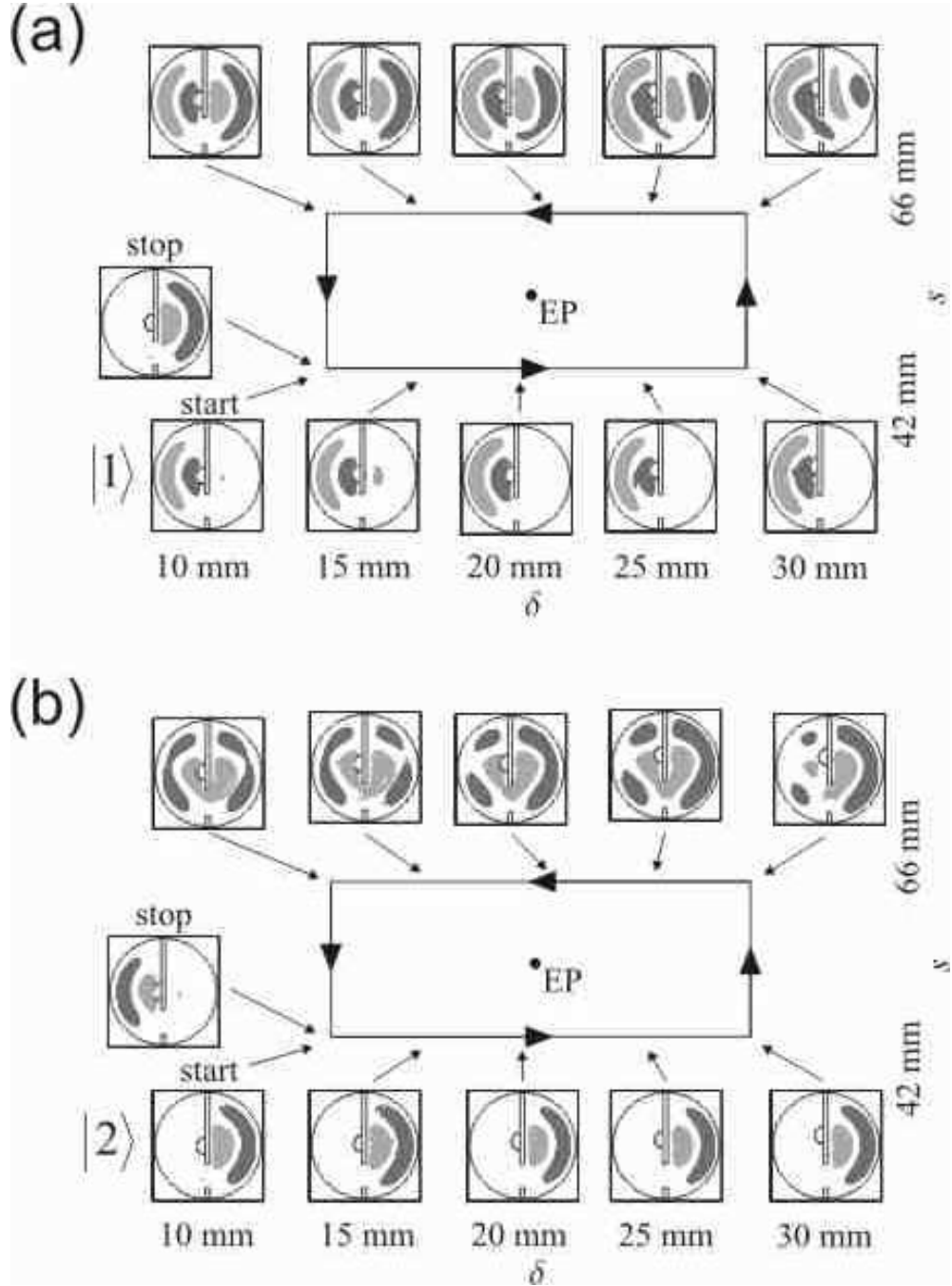}
\caption{Development of the reconstructed electrical field
distributions of two modes of the resonator shown in Fig.~\ref{fig:2}
while an EP is encircled. The initial states, i.e. the ''start''
configurations,  are labeled as $|1\rangle$ and $|2\rangle$ in
agreement with the definition (\ref{ExpBasisStates}). Their field
distributions can be reconstructed from the recorded nodal line
patterns for all settings $(s,\delta)$.} \label{fig:3} \end{figure}

The nodal line patterns of the electric field distributions could be
mapped by using a perturbation body method
\cite{DemboEP1,Sridhar,QCDTunnel}. Applying Berry's procedure
\cite{BerryGP} to these patterns we were able to reconstruct the
eigenfunctions of the cavity. Extending \cite{DemboEP1} we show here a
pair of modes, which for small couplings are localized in the adjacent
semi-circular halfs of the cavity (see Fig.~\ref{fig:3}).
Figure~\ref{fig:3} exhibits two reconstructed eigenfunctions of the
resonator for various parameter settings $(s,\delta)$. The two
shadings in Fig.~\ref{fig:3} can be associated with the two different
orientations of the electric field inside the microwave billiard
\cite{RichterBuch}. The wavefunctions of these modes can be mapped
out seperately even at a frequency crossing if the resonator is excited via
different antennas \cite{DemboEP2}. Following the theoretical analysis,
cf.~Eq.~(\ref{EigenvekAbkuerz}), we chose the wavefunctions for
$s=10$~mm and $\delta=42$~mm as basis states, i.e.
\begin{equation}
|1,2\rangle \equiv
|r_{1,2}\rangle\Big|_{(s=10~\mathrm{mm},\delta=42~\mathrm{mm})}\, .
\label{ExpBasisStates}
\end{equation}
The basis states are labeled as $|1\rangle$ in Fig.~\ref{fig:3}a and
$|2\rangle$ in Fig.~\ref{fig:3}b, respectively. They are chosen far
away from the EP, beforehand identified by studying the behavior of the
eigenvalues \cite{DemboEP1}, so that Eq.~(\ref{LargeB}) is fulfilled.
This implies that the basis states are approximately real
(cf.~Eq.~(\ref{EigenvecApprox})). At all  other parameter settings, the
eigenstates $|r_1\rangle$ and $|r_2\rangle$, are linear
combinations of $|1\rangle$ and $|2\rangle$, cf. \cite{DemboEP2}. Let
$\alpha$, $\beta$ be the expansion coefficients of $|r_1\rangle$ so
that
\begin{equation}
|r_1\rangle=\alpha|1\rangle+\beta|2\rangle\, .
\label{Psi1Entwicklung}
\end{equation}
The eigenstates are orthonormal in the biorthogonal sense, which
requires
\begin{equation}
|r_2\rangle=\beta|1\rangle-\alpha|2\rangle\, .
\label{Psi2Entwicklung}
\end{equation}
and
\begin{equation}
\alpha^2+\beta^2=1\, .
\end{equation}
The eigenfunctions remain approximately real for \emph{all} steps
displayed in Fig.~\ref{fig:3}, since a superposition of  $|1\rangle
,|2\rangle$ with complex expansion coefficients would have removed the
nodal lines while a superposition of the basis states with real
expansion coefficients simply shifts the nodal lines. The reason for
this is that the absolute value of the wave function of the complex
superposition is zero only where the coefficients of both
$|1\rangle$ and $|2\rangle$ vanish.

By varying $(s,\delta)$ in small steps, one EP has been encircled in
the $(s,\delta)$-plane. Both eigenfunctions were tracked continuously
during the sequence of eleven steps that form the closed loop around
the EP. The reconstructed wavefunctions clearly show that the basis
state $|1\rangle$ transforms to $|2\rangle$ (see Fig.~\ref{fig:3}a),
and that $|2\rangle$ transforms to $-|1\rangle$, which implies a
geometric phase of $\pi$. The data presented in  Fig.~\ref{fig:3}
therefore confirm in Berry's sense \cite{BerryGP,Lauber:94} the
appearance of a geometric phase which is picked up by
\emph{one} eigenvector when an EP is encircled:
 \begin{equation}
\left\{
\begin{array}{c}
|1\rangle \\
|2\rangle
\end{array}
\right\}
\circlearrowleft
\left\{
\begin{array}{c}
|2\rangle \\
-|1\rangle
\end{array}
\right\}
\label{ExpSingle}
\end{equation}
A recently suggested additional geometric phase \cite{Rotter}, which
also leads to mathematical inconsistencies \cite{Comment}, does not
appear.

\begin{figure}[ht]
\includegraphics[width=0.8\textwidth]{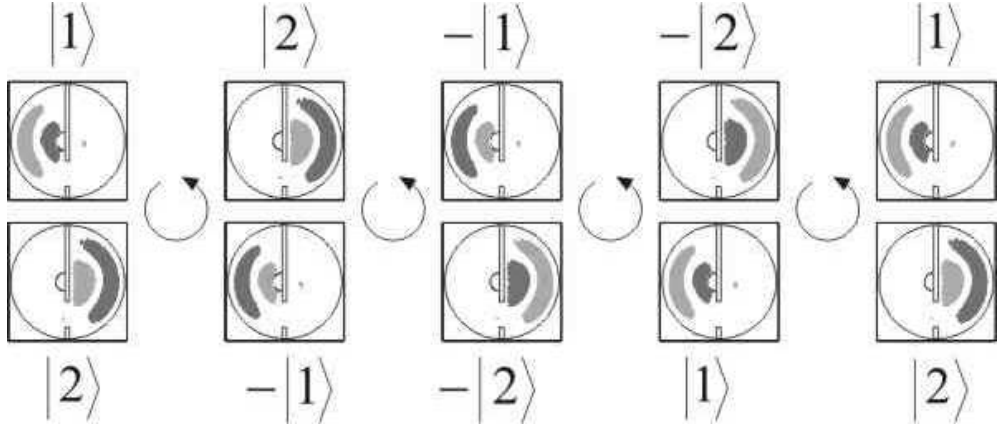}
\caption{Development of the basis states $|1\rangle$ and $|2\rangle$
during four consecutive turns around an EP. For each loop, symbolized by
$\circlearrowleft$, the development from the initial to the final states
has been tracked according to Fig.~\ref{fig:3}. Four turns are needed in
order to restore the original situation.}
\label{fig:4}
\end{figure}

The transformation scheme for four consecutive turns around a single EP
has been measured by repeatedly tracking the nodal line patterns along
the path shown in Fig.~\ref{fig:3}. The resulting geometric phases for
the basis states $|1\rangle$ and $|2\rangle$ can be derived from the
reconstructed wave functions shown in Fig.~\ref{fig:4}. Our experimental
rusults are in accordance
to the analytical result, Eq.~(\ref{PhasenA}),
see Fig.~\ref{fig:4}. The transformation scheme (\ref{PhasenA}) implies
that the eigenfunctions of the resonator have a branch point of fourth
root at the EP.

\section{Conclusion}
\label{sec:4}

We have shown analytically that the eigenfunctions of the two-state system
described by the
complex symmetric Hamiltonian modelling our microwave cavity experiment
transform according to Eqs.~(\ref{PhasenA}) or
(\ref{PhasenB}), when the parameters of the Hamiltonian are taken around
one EP, i.e. a fourth order branch point. The appearing geometric
phases are a  consequence of the normalization of the eigenfunctions
(\ref{EigenvecMod}). The eigenfunctions are approximately real on a
path encircling the EP, a property which is essential for their experimental
reconstruction according to
Berry \cite{BerryGP} from distributions of nodal lines mapped in
microwave cavity experiments.

We verified these results by performing an experiment with a normal
conducting microwave billiard consisting of two variably coupled
semi-circular resonators. The two modes we report on here, are for
small couplings localized in the adjacent semi-circular halfs of the
resonator. This allowed us to completely map their nodal line patterns
when the EP is encircled. The reconstructed wavefunctions confirm the
transformation schemes derived analytically. The experimental results
presented here show that the geometric phases occurring when an EP is
encircled agree with those observed in earlier experiments
\cite{DemboEP1} and with analytical and numerical calculations
\cite{EPTheo,BerryDennis,EPTheoNeu}. There is no experimental evidence
for any additional geometric phase factors \cite{Rotter}.

\section{Acknowledgment}
\label{sec:5}

We thank M.V.~Berry for pointing out his thoughts on branch points
which have been very helpful in finishing this manuscript.  This work
has been supported by DFG under contract number RI 242/16-3 and within
SFB 634 and the HMWK within the HWP.

\end{document}